\documentstyle[12pt]{article}

\catcode`@=11
\long\def\@caption#1[#2]#3{\par\addcontentsline{\csname
  ext@#1\endcsname}{#1}{\protect\numberline{\csname
  the#1\endcsname}{\ignorespaces #2}}\begingroup
    \small
    \@parboxrestore
    \@makecaption{\csname fnum@#1\endcsname}{\ignorespaces #3}\par
  \endgroup}
\catcode`@=12

\input{epsf}
\ifx\epsffile\undefined
\newlength{\epsfysize}
\def\epsffile#1#2#3#4]#5{}
\fi
\input{rotate}

\renewcommand{\theequation}{\thesection.\arabic{equation}}
\newcommand{\be}{\begin{equation}}   
\newcommand{\ee}{\end{equation}}
\newcommand{\bear}{\begin{eqnarray}}
\newcommand{\eear}{\end{eqnarray}}
\newcommand{\ba}{\begin{array}}      
\newcommand{\ea}{\end{array}}
\newcommand{\lae}{\begin{array}{c}\,\sim\vspace{-21pt}\\< \end{array}}
\newcommand{\gae}{\begin{array}{c}\,\sim\vspace{-21pt}\\> \end{array}}
\def\VEV#1{\left\langle #1\right\rangle}
\newcommand{\eL}{{\cal L}}
\newcommand{\Q}{{\cal Q}}
\newcommand{\R}{{\cal R}}
\newcommand{\m}{{\tilde m}}
\newcommand{\kol}{{\kappa/\lambda}}
\newcommand{\eol}{{\epsilon/\lambda}}
\newcommand{\kpol}{{\kappa'/\lambda}}
\begin{document}

\pagestyle{empty}
\begin{titlepage}
\def\thepage {}        

\title{\bf Generic and Chiral Extensions \\ [2mm] of the
Supersymmetric Standard Model\\ [1cm]}

\author{ {\small \bf Hsin-Chia Cheng,  Bogdan A.~Dobrescu and
Konstantin T.~Matchev} \\  \\ {\small {\it Theoretical Physics
Department}}\\   {\small {\it Fermi National Accelerator
Laboratory}}\\ {\small {\it Batavia, Illinois, 60510, USA
\thanks{e-mail addresses: hcheng@fnal.gov, bdob@fnal.gov,
matchev@fnal.gov} }}\\ }

\date{ }

\maketitle

   \vspace*{-9.4cm}
\noindent

\makebox[10.0cm][l]{November 12, 1998} FERMILAB-PUB-98/357-T \\ [1mm]

\vspace*{8.5cm}

\baselineskip=18pt

\begin{abstract}

{\normalsize We construct extensions of the Standard Model in which
the gauge symmetries  and supersymmetry prevent the dangerously large
effects that may potentially  be induced in a supersymmetric standard
model by Planck scale physics.  These include baryon number violation,
flavor changing neutral currents, the $\mu$ term, and masses for
singlet or vector-like fields under the Standard Model gauge group.
For this purpose we introduce an extra non-anomalous $U(1)_\mu$ gauge
group.  Dynamical supersymmetry breaking in a secluded sector triggers
the breaking of the $U(1)_\mu$ and generates soft masses for the
superpartners via gauge mediation, with the scalars possibly receiving
sizable contributions from the $U(1)_\mu$ D-term.  We find several
classes of complete and calculable models, in which the messengers do
not present cosmological problems and neutrino masses can also be
accomodated.  We derive the sparticle spectrum in these models and
study the phenomenological consequences. We give an exhaustive list of
the potential experimental signatures and discuss their observability
in the upcoming Tevatron runs. One class of models exhibits
interesting new discovery channels, namely $WW\not\!\!\!E_T$,
$W\gamma\not\!\!\!E_T$ and $WZ\not\!\!\!E_T$, which arise when the
next-to-lightest supersymmetric particle is a short-lived $SU(2)_W$
neutralino. }

\end{abstract}

\vfill
\end{titlepage}

\baselineskip=18pt
\pagestyle{plain} \setcounter{page}{1}

\section{Introduction}
\label{introduction}
\setcounter{equation}{0}

The gauge structure of the Standard Model (SM) has two important
consequences in agreement with the experimental data: the proton
stability is ensured by a discrete  baryon number symmetry, and there
are no tree-level flavor changing  neutral currents (FCNC). On the
other hand, a mass term for the Higgs doublet is not prevented by any
symmetry in the Standard Model, leading to the well known  hierarchy
and naturalness problems.

The Minimal Supersymmetric Standard Model (MSSM) solves the
naturalness  problem (because supersymmetry [SUSY] ensures the
cancellation of  the quadratic divergences in the Higgs self-energy),
but does not solve the hierarchy problem ({\it i.e.}, it does not
explain the large hierarchy between the $\mu$-term or the soft
supersymmetry breaking parameters and the Planck scale). Moreover, the
MSSM does not have the attractive features of the Standard Model
mentioned above: the gauge structure  allows both proton decay
operators and FCNCs.

The resolution to these issues may be provided by physics beyond the
MSSM.  For example, the exponential hierarchy between the soft
breaking  parameters and the Planck scale is naturally produced if
supersymmetry is dynamically broken.  The tree level FCNCs are
eliminated if there is a global $R$-symmetry, while radiative FCNCs
can be kept sufficiently  small if supersymmetry breaking is
communicated to the MSSM fields by generation-independent  gauge
interactions.  The proton decay operators can be avoided by invoking a
discrete baryon number symmetry, and the $\mu$-term can be kept small
compared with  the Planck scale by a discrete symmetry whose breaking
is triggered by  the supersymmetry breaking. Likewise, some discrete
symmetries may be used to eliminate other unacceptable operators
associated with the new physics beyond the MSSM, such as large mass
terms for  the gauge singlets required by many gauge mediated
supersymmetry breaking models.

At present, all viable supersymmetric extensions of the Standard Model
rely on the existence of some discrete symmetries which are not known
to be associated with gauge symmetries. This situation is rather
unfortunate  given that currently it is not known whether the global
symmetries are  preserved by quantum gravitational effects.  In fact
there are some arguments that support the idea that any global
symmetry is badly violated in the presence of nonperturbative
gravitational effects \cite{wormhole}:  the global charges may leak
through a wormhole, or they may disappear  into a black hole which may
evaporate. In the low energy effective theory,  these effects give
rise to gauge-invariant operators which break explicitly the global
symmetries.  Generically, one expects these operators to have
coefficients of order one times the appropriate power of the Planck
scale.  This results in a $\mu$ term which is too large by 16 orders
of magnitude, dimension-four baryon number violating operators which
have  coefficients 22 orders of magnitude larger than the upper bound
set by the proton decay measurements, and other disastrous effects.
However, in certain cases where general relativity is modified at
energies significantly below the Planck scale \cite{relativ}, it is
possible to suppress the coefficients of the symmetry violating
operators.  In any case, the extent of global symmetry violation
appears to be  highly sensitive to the underlying theory of quantum
gravity, which  is not known yet.

Hence, it would be useful to show that the global symmetries required
in the MSSM are remnants of some spontaneously broken gauge
symmetries. In string theory and M-theory there are situations where
discrete symmetries in the low energy theory are remnants of gauge
groups spontaneously broken by the string dynamics \cite{string}.
However, it is by no means clear that once the appropriate vacuum of a
viable string theory is found, the necessary discrete symmetries of
the MSSM would be preserved.  Therefore, it has been often attempted
to extend the SM gauge group  so that the harmful operators allowed in
the MSSM are no longer gauge invariant. The simplest extension is to
include a spontaneously broken $U(1)$ gauge symmetry, and it has been
used to avoid baryon number violating operators \cite{Weinberg} or a
large $\mu$-term \cite{mu}. Nevertheless, no chiral ({\it i.e.}
without gauge invariant mass terms)  and generic ({\it i.e.} without
unnaturally small dimensionless couplings) supersymmetric model has
been constructed yet.

In a previous paper \cite{CDM} we showed that a new $U(1)$ gauge
symmetry, in conjunction with supersymmetry and the standard  $SU(3)_C
\times SU(2)_W \times U(1)_Y$ gauge group, is  sufficient to prevent
{\it any} mass terms (including the $\mu$-term),  so that the only
fundamental dimensional parameter is the Planck scale.  Although this
is a chiral supersymmetric model,  it relies as much as the MSSM on
discrete symmetries to eliminate the proton decay operators.  Given
that our goal is to construct a self-consistent  theory which does not
invoke arbitrary assumptions about quantum gravity, we must use a
gauge symmetry to eliminate the proton decay operators, as  well as
any other dimension-four and higher operators forbidden by
phenomenology.

In this paper we show that the gauge group introduced in \cite{CDM} is
in fact sufficient to replace any discrete symmetry required by the
phenomenological constraints, provided the charge assignments under
the new $U(1)$ gauge symmetry are chosen carefully.  We find several
classes of phenomenologically viable models of this type.  These are
chiral and generic supersymmetric models to the extent that we do not
attempt to explain the quark and lepton masses, so that we allow
Yukawa couplings as small as $\sim \lambda_e \sim 10^{-5}$.

An interesting feature of our models is that the new $U(1)$
communicates supersymmetry breaking from a dynamical supersymmetry
breaking (DSB) sector to the MSSM fields.  Furthermore, unlike the
previous  models in which a spontaneously broken $U(1)$ mediates
supersymmetry breaking \cite{Mohapatra, bdob}, the existence of a DSB
sector and of a sector responsible for gaugino masses are required by
the gauge anomaly cancellation conditions.  As a consequence, the
superpartner spectrum is quite distinctive.  We discuss the resulting
phenomenology and find some interesting cases with unexpected
experimental signatures.

The plan of the paper is as follows.  In Section \ref{framework} we
discuss the theoretical and phenomenological constraints, and use them
to find a fairly exhaustive class of viable models.  In Section
\ref{phenomenology} we study the phenomenology of this class of
models.  We describe their low-energy spectrum and discuss the
experimental search strategy in each of the typical scenarios, by
singling out the most promising channels to look for in the upcoming
Tevatron runs.  The implications of relaxing some of the
phenomenological constraints are considered in Section
\ref{conclusions}, where we also draw our  conclusions.

\section{Framework and Constraints}
\label{framework}
\setcounter{equation}{0}

If the gauge group acting on the MSSM chiral superfields is  $SU(3)_C
\times SU(2)_W \times U(1)_Y \times U(1)_\mu$,   then the $H_u H_d$
term in the  superpotential is forbidden provided the  $U(1)_\mu$
charges of  the two Higgs superfields satisfy $z_{H_u} + z_{H_d} \neq
0$.  In order to produce a Higgsino mass, we introduce a $S H_u H_d$
term in the  superpotential, where the Standard Model singlet $S$ has
$U(1)_\mu$ charge  $z_S = - z_{H_u} - z_{H_d}$, and its scalar
component acquires a vev.

In order to have quark and lepton masses and mixings (we allow lepton
mixings in compliance with the recent Super-Kamiokande results
\cite{superK}), the most general Yukawa couplings of the Higgs
doublets to quarks and leptons require the $U(1)_\mu$ charges of the
quark and lepton superfields,  $Q_i,\bar{U}_i, \bar{D}_i, L_i,
\bar{E}_i, {\nu_R}_i$ ($i = 1, 2, 3$ is a generational index), to be
family-independent and to satisfy  \bear  z_Q &=& -z_{H_u} -
z_{\bar{U}} = -z_{H_d} - z_{\bar{D}} ~,  \nonumber \\ z_L &=& -z_{H_u}
- z_{\nu} = -z_{H_d} -z_{\bar{E}}.   \eear  These conditions can be
relaxed if the quark and lepton  mass matrices have textures produced
by a non-standard mechanism, such as Frogatt-Nielsen~\cite{FN}, but we
will not study this possibility here.

For $U(1)_\mu$ to be anomaly free, additional chiral superfields have
to be included. The $[SU(3)_C]^2 \times U(1)_\mu$, $[SU(2)_W]^2 \times
U(1)_\mu$,  $[U(1)_Y]^2 \times U(1)_\mu$ anomalies from the MSSM
fields  are\footnote{We use the normalization ${\rm tr}(T^c \{T^a,
T^b\})$ for the anomalies, so the  $[U(1)_Y]^2 \times U(1)_\mu$
anomaly from a field with $U(1)_Y \times U(1)_\mu$ charges $(y, \, z)$
is $2y^2 z$.}  \bear
\label{A3}
(A3) & \equiv & \left[SU(3)_C\right]^2 \times U(1)_\mu :\quad 3\left(2
z_Q + z_{\bar{U}} + z_{\bar{D}}\right) =  3 z_S , \nonumber \\
\label{A2}
(A2) &\equiv &\left[SU(2)_W\right]^2 \times U(1)_\mu : \quad
9 z_Q+3z_L-z_S , \nonumber \\
\label{A1}
(A1) &\equiv &\left[U(1)_Y\right]^2 \times U(1)_\mu :\quad
-9 z_Q-3z_L+7z_S .  \eear They have to be cancelled by fields which
carry both SM and $U(1)_\mu$ quantum numbers. In order not to
introduce anomalies to the SM gauge group, and to be able to decouple
at low energies after $U(1)_\mu$ is broken, these fields should be
vector-like under the SM gauge group. As a result, they can naturally
be identified with the messengers of gauge mediated supersymmetry
breaking.

The masses of the messengers are induced by a $X \phi \bar{\phi}$ term
in the superpotential, where $\phi$, $\bar{\phi}$ represent the
messenger fields, $X$ is a SM singlet, and their $U(1)_\mu$ charges
are related by $z_\phi +z_{\bar{\phi}}=-z_X$.

In order to generate the soft supersymmetry breaking masses for the
MSSM fields through gauge mediation, the $X$ superfield should have
both scalar and $F$-type vevs with $\VEV{F_X}/\VEV{X} \sim 10^4-10^5$
GeV and hence can not be identified with the $S$ field (otherwise it
will give a too big $B$ term for the Higgs sector). The simplest way
to have a (local) minimum in which $S$ and $X$ obtain the desired vevs
is having only one $X$ field which couples to all messengers, and
introducing another SM singlet $N$, with the superpotential in
Ref.~\cite{CDM}, \be
\label{WSXN}
W= f X \phi \bar{\phi} + \frac{\lambda}{2} X N^2 -  \frac{\epsilon}{2}
S N^2 + \kappa S H_u H_d \, .  \ee Phenomenological contraints require
$\lambda^{3/2} < \epsilon \ll \lambda \ll f \sim 1$ \cite{CDM}. For
$\kappa > \sqrt{\lambda^2 +\epsilon^2}$, there is a desired minimum in
which  all $S$, $X$, and $N$ fields obtain vevs, after they receive
negative masses squared of their scalar components from the DSB sector
\cite{CDM}.  This choice of superpotential imposes the following
relation between the $U(1)_\mu$ charges of $S,\, X,$ and $N$ fields \be
\label{ZSXN}
z_S =z_X=-2 z_N.  \ee There are two other possible terms in the
superpotential, which are allowed by the gauge symmetries, $f' S \phi
\bar{\phi}$ and $\kappa' X H_u H_d$.  The minimum will not be affected
if the $\kappa'$ coupling is small. The first term  contributes to the
messenger masses and the second term gives extra contribution to the
$B$ term of the Higgs sector. We assume that the couplings $f'$ and
$\kappa'$,  if they exist, are small so that the messenger masses
receive dominant contributions from $X$ and the desired minimum is not
destablized.

For a vector-like pair of messengers $\phi, \bar{\phi}$ with $SU(3)_C$
index $T_{3\phi}=(T_{3\bar{\phi}})$, (normalized to 1/2 for the
fundamental representation), $SU(2)_W$ index $T_{2\phi}$, and $U(1)_Y$
charges $\pm y_\phi$, the contributions to the anomalies (A3)--(A1)
are \bear &(A3)& \quad 2T_{3\phi}(z_\phi+z_{\bar{\phi}}) = -2T_{3\phi}
z_X,  \nonumber \\ &(A2)& \quad 2T_{2\phi}(z_\phi+z_{\bar{\phi}}) =
-2T_{2\phi} z_X,  \nonumber \\ &(A1)& \quad
2y_\phi^2(z_\phi+z_{\bar{\phi}}) = -2y_\phi^2 z_X .  \eear A messenger
field, $a$, which is real under SM with $U(1)_\mu$ charge $-z_X/2$ can
obtain its mass from the vev of $X$ without its conjugate partner. In
this case, its contributions to (A3)--(A1) are $-T_{3a} z_X$, $-T_{2a}
z_X$, and $-y_a^2 z_X$, respectively.

To cancel the anomalies coming from the MSSM sector [eq.~(\ref{A1})],
the messengers have to satisfy \bear
\label{cond3}
3z_S - \sum_{r_3} T_{3r_3} z_S =0, \\
\label{cond2}
9z_Q +3z_L-z_S-\sum_{r_2} T_{2r_2} z_S =0, \\
\label{cond1}
-9z_Q-3z_L+7z_S-\sum_{r_1} y_{r_1}^2 z_S =0,  \eear where $r_i$ runs
over all messenger representations  (counting the SM vector-like pair
separately) under  $SU(3)_C$, $SU(2)_W$, and $U(1)_Y$ respectively.
The gauge mediated contributions to the soft masses of the MSSM fields
transforming under  $SU(3)_C$, $SU(2)_W$, and $U(1)_Y$ are
proportional to the messenger multiplicity factors  \be \Delta \beta_3
\equiv \sum_{r_3} T_{3r_3}, \quad \Delta \beta_2 \equiv \sum_{r_2}
T_{2r_2}, \quad \Delta \beta_1 \equiv \sum_{r_1} y_{r_1}^2, \ee which
are just the changes of the one-loop $\beta$-function coefficients of
the corresponding gauge groups due to the messenger fields. From
eq.~(\ref{cond3}) we see that \be
\label{b3}
\Delta \beta_3 = \sum_{r_3} T_{3r_3}=3 , \ee which means the messenger
sector should either contain three pairs of {\bf 3} and $\bf \bar{3}$,
or one {\bf 8} under $SU(3)_C$.  Combining eqs.~(\ref{cond2}) and
(\ref{cond1}) we obtain another constraint on the messenger sector, \be
\label{b2plusb1}
\Delta \beta_2+\Delta \beta_1 = \sum_{r_2} T_{2r_2}  +\sum_{r_1}
y_{r_1}^2 =6, \ee which limits $\Delta \beta_2$ and $\Delta \beta_1$
to a discrete set of choices.

The only possible messengers which can satisfy eqs.~(\ref{b3})  and
(\ref{b2plusb1}) (and do not cause the SM gauge couplings blowing up
below the Planck scale) are the ones transforming  under $SU(3)_C
\times SU(2)_W$ as ({\bf 3, 2}), ({\bf 3,1}), ({\bf 8,1}), ({\bf
1,2}), ({\bf 1,3}),  ({\bf 1,1}) and their conjugates. If they have
arbitrary hypercharges, then in general they can not decay into MSSM
fields. They will be stable and form bound states with fractional
electric charges, which may cause cosmological problems unless a late
period of inflation is incorporated.  To avoid that, the hypercharges
of the messenger fields are fixed up to additive integers by the
hypercharges of the MSSM fields.  Imposing the conditions (\ref{b3})
and (\ref{b2plusb1}), we find that the messenger sector can only
consist of fields among $q=({\bf 3, 2}, +1/6)$, $\bar{u}=({\bf
\bar{3}, 1}, -2/3)$, $\bar{d}= ({\bf \bar{3}, 1}, +1/3)$, $a=({\bf
8,1}, 0)$, $l=({\bf 1, 2}, -1/2)$,  $w=({\bf 1, 3}, 0)$,
$\bar{e}=({\bf 1, 1}, +1)$, and their conjugates.  There are 16
possible combinations with four different sets of $(\Delta \beta_3,\;
\Delta \beta_2,\; \Delta \beta_1)$, which  are shown in
Table~\ref{TableMess}.
\begin{table}[t]
\centering \renewcommand{\arraystretch}{1.5}
\begin{tabular}{|c| |c|c|c|c|c|c|c| |c|c|c|}\hline
Model  & $d\overline{d}$ & $u\overline{u}$ & $q\overline{q}$ & $a$ &
 $l\overline{l}$ & $w$ & $e\overline{e}$ & $\Delta\beta_3$  &
 $\Delta\beta_2$ & $\frac{3}{5}\Delta\beta_1$ \\ \hline \hline 1a &3
 &--  &-- &--  &1 &-- &1  &3 &1 &3 \\ \hline 1b &2 &1  &-- &--  &1 &--
 &--  &3 &1 &3 \\ \hline 1c &-- &--  &-- &1  &1 &-- &2  &3 &1 &3 \\
 \hline \hline 2a &3 &-- &-- &-- &2 &-- &-- &3 &2 &2.4 \\ \hline 2b &3
 &--  &-- &--  &-- &1  &1  &3 &2 &2.4 \\ \hline 2c &2 &1  &-- &--  &--
 &1  &--  &3 &2 &2.4 \\ \hline 2d &-- &--  &-- &1 &2 &-- &1  &3 &2
 &2.4 \\ \hline 2e &-- &--  &-- &1  &-- &1  &2  &3 &2 &2.4 \\
 \hline\hline 3a &3 &--  &-- &--  &1 &1  &--  &3 &3 &1.8 \\ \hline 3b
 &1 &--  &1 &--  &-- &--  &1  &3 &3 &1.8 \\ \hline 3c &-- &1  &1 &--
 &-- &--  &--  &3 &3 &1.8 \\ \hline 3d &-- &--  &-- &1  &3 &--  &-- &3
 &3 &1.8 \\ \hline\hline 4a &3 &--  &-- &--  &-- &2  &--  &3 &4 &1.2
 \\ \hline 4b &1 &--  &1 &--  &1 &--  &--  &3 &4 &1.2 \\ \hline 4c &--
 &--  &-- &1  &2 &1  &--  &3 &4 &1.2 \\ \hline 4d &-- &--  &-- &1  &--
 &2  &1  &3 &4 &1.2 \\ \hline
\end{tabular}
\parbox{5.5in}{
\caption{ Possible number of messenger representations, and the
corresponding contributions to the gauge coupling beta functions.  The
factor of 3/5 in front of $\Delta\beta_1$ corresponds to the $SU(5)$
normalization of the hypercharge.
\label{TableMess}}}
\end{table}

Already from the above simple constraints, we can see that there are
only four possible combinations of the gauge mediated contributions to
the soft masses of the MSSM fields. In particular for the SM gaugino
masses, which only receive masses from gauge mediation, their ratios
are fixed to these four cases, independent of the assumption that
there are no states with fractional electric charges.  If the
$U(1)_\mu$ $D$ term and the other contributions are small compared to
the gauge mediated contributions, then the complete sparticle spectrum
is determined to a large extent in these four cases. For larger
$U(1)_\mu$ $D$ term contributions, we also need to know the specific
$U(1)_\mu$ charges of the MSSM fields, in order to predict the scalar
superpartner spectrum.

In addition to (\ref{cond3})--(\ref{cond1}), the $U(1)_\mu$ charges
also have to satisfy the $U(1)_Y\times [U(1)_\mu]^2$, $U(1)_\mu$,  and
$[U(1)_\mu]^3$ anomaly cancellation conditions.  In general, the
latter two anomalies are not cancelled by the combination of the MSSM
and messenger sector. Therefore, some fields from the DSB  sector have
to carry $U(1)_\mu$ charges, so that $U(1)_\mu$ can communicate
supersymmetry breaking to both the messenger sector and to the MSSM
chiral superfields.  It is remarkable that the existence of the three
sectors (MSSM, messenger  and DSB) is required by the mathematical
consistency of the theory  (namely the anomaly cancellation
conditions). We consider this situation an improvement compared with
the original gauge-mediated models \cite{DNS,dnns} in which the three
different sectors are introduced only for  phenomenological reasons.

If the DSB sector dynamics does not break $U(1)_\mu$, then its
contributions to the $U(1)_\mu$ and  $[U(1)_\mu]^3$ anomalies can be
represented by low energy effective composite degrees of freedom {\it
a la} 't~Hooft~\cite{tHooft}.  The simplest example is the 3-2 model
\cite{ADS,DNS}, where after $SU(3)$ becomes  strong and breaks
supersymmetry, there is one light field charged under the unbroken
``messenger'' $U(1)$. Other DSB models have a different number of
light composite fields with various $U(1)_\mu$ charge ratios. For
simplicity, in searching  for solutions, we restrict ourselves to the
cases with no more than 2 extra such SM neutral and $U(1)_\mu$ charged
composite fields from the DSB sector. A renormalizable and calculable
example  of a DSB model which gives rise to two light $U(1)_\mu$
charged composite fields is the $SU(4)\times SU(3)$ model
\cite{PST,AMM,CDM}. A brief description of the model and its
$U(1)_\mu$ charge assignments  is presented in Appendix~A.

There are several additional constraints we impose when we search for
models. We allow the right-handed neutrinos to acquire Majorana
masses, so the $U(1)_\mu$ charges of the right-handed neutrinos have
to be $z_\nu=-z_S/2$ or $-z_N/2$ if they receive masses from $S$ or
$N$ vevs. Note that we avoid $z_\nu=0$ because in that case the field
content would not be chiral: the right-handed neutrinos would be gauge
singlets, and a Planck scale mass for $L_i$ and $H_u$ would be
potentially induced.  For $z_\nu =-z_S/2 (=z_N)$, the operators $N L_i
H_u$ are gauge invariant, and give rise to the bilinear $R$  parity
violating terms after $N$ acquires a vev.  The phenomenological
constraints on these bilinear terms (e.g., from flavor changing
neutral currents) require the  couplings of the $N L_i H_u$
interactions to  be very small. We will therefore only concentrate on
the case $z_\nu =-z_N/2$. In this case we will find that $R$ parity
conservation is an automatic consequence of the gauge  symmetry.

We are free to choose $z_S > 0$ (note that $z_S \neq 0$ to avoid  a
large $\mu$ term), which implies that the $U(1)_\mu$ $D$ term is
positive. We will require the $U(1)_\mu$ charges for ordinary quarks
and leptons to be non-negative, so that they do not receive negative
masses squared from the  $U(1)_\mu$ $D$ term. This may not be
necessary if the positive contributions from gauge mediation are
larger than the negative $D$ term contributions. However, the squarks
and sleptons receive $D$ term masses at a higher scale, so the SM
gauge group may be broken before the gauge mediated contributions can
turn on.  Therefore, we do not search for models with negative quark
or lepton charges.

Finally, if the messenger fields do not couple to the MSSM fields,
they are stable. For typical values of the messenger masses, they will
overclose the universe \cite{DGP}, unless diluted by a late period of
inflation. We therefore require that the $U(1)_\mu$ charges allow the
messenger fields to couple to the MSSM fields so that the messenger
fields can decay into MSSM fields before nucleosynthesis. This
requires the relevant matter-messenger couplings to be suppressed by
no more than one power of the Planck mass \cite{DGP}.  At the same
time, the matter-messenger interactions which can  induce too fast
proton decays should be forbidden (including the lepton number
conserving decays to gravitinos~\cite{Choi}).

The $U(1)_\mu$ charges of the MSSM fields can be expressed in terms of
the 4 charges, $z_Q$,  $z_L$, $z_{H_u}$, and $z_S$, from the
requirements of the MSSM superpotential interactions.  The Majorana
masses of the right-handed neutrinos impose a relation among $z_L$,
$z_{H_u}$, and $z_S$ ($-z_L-z_{H_u}=z_\nu=z_S/4$).  Among the anomaly
conditions (\ref{cond3})--(\ref{cond1}), only 2 combinations have been
used. The other one, which can be taken as (\ref{cond2}), gives
another constraint among $z_Q$, $z_L$, and $z_S$ for each choice of
$\Delta \beta_2$, \be 9z_Q +3z_L -(1+\Delta\beta_2)z_S=0\, .  \ee We
choose the overall charge normalization by fixing $z_S$.  The
$U(1)_\mu$ charges of the MSSM fields then depend only on  one
independent charge, for example $z_Q$, and its range is limited by the
requirement that the quark and lepton $U(1)_\mu$ charges are
non-negative. The $U(1)_\mu$ charges of the MSSM fields as a function
of $z_Q$, and the allowed range for $z_Q$ for each case of
$\Delta\beta_2$ are shown in Tables~\ref{MSSMcharges} and
\ref{zQrange}, respectively.
\begin{table}[h]
\renewcommand{\arraystretch}{1.8} \centering
\begin{tabular}{|c|c|}\hline
$Q_i$ &  $z_Q$    \\ \hline $\bar{U}_i$ & $-4z_Q+5 +4
\left(\Delta\beta_2-2\right)/3$ \\ \hline $\bar{D}_i$ & $ 2z_Q-1 -4
\left(\Delta\beta_2-2\right)/3$ \\ \hline $L_i$ & $-3z_Q+4 +4
\left(\Delta\beta_2-2\right)/3$ \\ \hline $\bar{E}_i$ & $ 6z_Q-5 -8
\left(\Delta\beta_2-2\right)/3$ \\ \hline $\nu_i$ & $1$ \\ \hline
$H_u$ & $ 3z_Q-5 -4 \left(\Delta\beta_2-2\right)/3$ \\ \hline $H_d$ &
$-3z_Q+1 +4 \left(\Delta\beta_2-2\right)/3$ \\ \hline
\end{tabular}
\caption{$U(1)_\mu$ charges of the MSSM fields in terms of $z_Q$, with
the normalization $z_S=4$.}
\label{MSSMcharges}
\end{table}
\begin{table}[ht]
\centering \renewcommand{\arraystretch}{1.8}
\begin{tabular}{|c|ccccc|} \hline
$\Delta\beta_2=1$ & $14/36$ & $\leq$ & $z_Q$ & $\leq$ & $32/36$ \\
\hline $\Delta\beta_2=2$ & $30/36$ & $\leq$ & $z_Q$ & $\leq$ & $45/36$
\\ \hline $\Delta\beta_2=3$ & $46/36$ & $\leq$ & $z_Q$ & $\leq$ &
$57/36$ \\ \hline $\Delta\beta_2=4$ & $66/36$ & $\leq$ & $z_Q$ &
$\leq$ & $69/36$ \\ \hline
\end{tabular}
\caption{The range of $z_Q$ for all MSSM quark and lepton charges
being non-negative, normalizing to $z_S=4$.}
\label{zQrange}
\end{table}

For the cases in Table~\ref{TableMess} and ``reasonably simple''
$U(1)_\mu$ charges in the corresponding allowed range, we  search
numerically for the messenger and (DSB sector composite) singlet
charges which satisfy the rest of the anomaly constraints,  allow
messengers to decay fast enough, and forbid too rapid proton
decay. Some of the solutions satisfying all the constraints are listed
in Table~\ref{Tablesolutions}.
\begin{table}[tp]
\centering \renewcommand{\arraystretch}{1.5}
\begin{tabular}{||c||c|r|r|r|r||} \hline\hline 
Fields      & \multicolumn{5}{c||}{Models}         \\ \cline{2-6}  &
            1a(i) & 2a(i) & 2a(ii)& 2a(iii)& 2a(iv)\\ \hline\hline $Q$
            &   $2$ &   $8$ &  $ 1$ & $  10$ & $ 11$ \\ \hline
            $\bar{U}$   &   $3$ &  $13$ &  $ 1$ & $   5$ & $  1$ \\
            \hline $\bar{D}$   &   $5$ &   $7$ &  $ 1$ & $  11$ & $
            13$ \\ \hline $L$         &   $2$ &  $12$ &  $ 1$ & $   6$
            & $  3$ \\ \hline $\bar{E}$   &   $5$ &   $3$ &  $ 1$ & $
            15$ & $ 21$ \\ \hline $\nu_R$     &   $3$ &  $ 9$ &  $ 1$
            & $   9$ & $  9$ \\ \hline $H_u$       &  $-5$ & $-21$ &
            $-2$ & $ -15$ & $-12$ \\ \hline $H_d$       &  $-7$ &
            $-15$ &  $-2$ & $ -21$ & $-24$ \\ \hline \hline $S,X$ &
            $12$ &  $36$ &  $ 4$ & $  36$ & $ 36$ \\ \hline $N$ &
            $-6$ & $-18$ &  $-2$ & $ -18$ & $-18$ \\ \hline
            $\bar{d}_i$ &  $-4$ & $ -2$ &  $ 0$ & $   2$ & $  4$ \\
            \hline $d_i$       &  $-8$ & $-34$ &  $-4$ & $ -38$ &
            $-40$ \\ \hline $l_i$       &  $-1$ &  $12$ &  $ 1$ & $ 6$
            & $  3$ \\ \hline $\bar{l}_i$ & $-11$ & $-48$ &  $-5$ & $
            -42$ & $-39$ \\ \hline $\bar{e}$   &  $-4$ &  --   &  -- &
            --   &  --   \\ \hline $e$         &  $-8$ &  --   & --
            &   --   &  --   \\ \hline $b_1$       & $-12$ & $-39$ &
            $-4$ & $ -36$ & $-36$ \\ \hline $b_2$       & $18$ &  $90$
            &  $10$ & $  90$ & $ 90$ \\ \hline\hline & $QL\bar{d},\,
            \bar{U}\bar{E}d, $  & \multicolumn{4}{c||}{$\bar{E} H_d
            l,\; \nu_R H_u ,l$} \\ Messenger   & $ \bar{D}\nu_R d, \,
            LL\bar{e},$  & \multicolumn{4}{c||}{$X L \bar{l},$} \\
            decay   & $\bar{E}\nu_R e,\, NQ\bar{D}l, $  &
            \multicolumn{4}{c||}{$N Q L \bar{d},$} \\ operators &
            $NL\bar{E}l,\, \bar{E} \nu_R H_d l,$  &
            \multicolumn{4}{c||}{$N \bar{U} \bar{D} \bar{d},$} \\ & $
            XN H_u l,\, \nu_R \nu_R H_u l$  & \multicolumn{4}{c||}{$Q
            \nu_R H_d \bar{d}$} \\ \hline
\end{tabular}
\caption{Solutions for the $U(1)_\mu$ charges (normalized to
integers), which satisfy all the constraints. In models 1a(i) and
2a(ii) we find many other possible solutions with different messenger
charges, including different charges for the different $d_i$'s and
$l_i$'s. Here we only list one example for each case.}
\label{Tablesolutions}
\end{table}
The fields $b_{1,2}$ are the light composite superfields from the  DSB
sector which carry $U(1)_\mu$ charges. Note that mass terms  involving
$b_{1,2}$ and $S$ or $X$ can be generated only by higher  dimensional
operators involving the fundamental fields from the DSB, and therefore
are Planck-scale suppressed.  We find solutions in only a few out of
the  16 cases because of the restriction that there are no more than
two $b_i$ superfields. If we relax this simplifying assumption  and
allow more singlets,  there could be solutions in other cases as well.

The low energy MSSM spectrum and phenomenology depend mainly on
$\Delta\beta_{1, 2, 3}$ and the $U(1)_\mu$ charges of the MSSM fields.
They have little dependence on the exact compositions and charge
assignments of the mesenger and DSB sectors as long as the mixings
between the MSSM fields and messenger fields are small. We will
discuss the phenomenology in the next section.

\section{Phenomenology}
\label{phenomenology}
\setcounter{equation}{0}

\subsection{Particle spectrum}

First we shall briefly review the parameter space of this class of
models\footnote{For more details, we refer the reader to
Ref.~\cite{CDM}.}  and discuss the possible particle spectra arising
in each case.  For the rest of this section, we shall use the
$U(1)_\mu$ charge normalization $z_S=4$ and rescale the charges in
Table~\ref{Tablesolutions} correspondingly.

The desired minimum of the potential is at  $\langle H_u \rangle =
\langle H_d \rangle = 0$ (at the scale of $U(1)_\mu$ breaking) and \be
\langle N^2 \rangle ={24\m^2\over\lambda^2+\epsilon^2},  \quad \langle
X \rangle = {\epsilon\over\lambda}\langle S \rangle, \quad \langle S^2
\rangle = {\lambda^2 \over \lambda^2+\epsilon^2} \left( {\xi^2 \over4}
+{ \m^2\over g_\mu^2} +{12\m^2\over\lambda^2+\epsilon^2} \right) ~.
\label{vevs}
\ee The corresponding SUSY-breaking $F$ and $D$-terms are induced at
the $U(1)_\mu$ breaking scale \be M_\mu\equiv g_\mu \langle N \rangle
\simeq  2\sqrt{6} {g_\mu\over \lambda}\tilde m \quad (\gg \m),
\label{Mmu}
\ee where $g_\mu$ is the $U(1)_\mu$ gauge coupling, and are given by
\be \langle F_N \rangle =0, \quad \langle F_X \rangle =
\frac{\lambda}{2} \langle N^2 \rangle \simeq \sqrt{6} \m \langle N
\rangle, \quad \langle F_S \rangle = -\frac{\epsilon}{2}  \langle N^2
\rangle, \quad g_\mu^2 \langle D \rangle = 4 \m^2.
\label{F}
\ee

The $\langle X \rangle$ and $\langle F_X \rangle$ vevs provide the
SUSY preserving and breaking masses for the messenger fields $\phi$
and $\bar{\phi}$.  The gauge singlets $X$, $S$ and $N$ also get
masses.  Their fermionic components mix with the $U(1)_\mu$ gaugino to
form two Dirac fermions, with masses $\sim24(g_\mu/\lambda)\tilde m$
and $\sim4\tilde m$, respectively.  The scalar components of the
singlets also mix, and the resulting mass spectrum consists of a
massless Nambu-Goldstone boson, eaten by the $U(1)_\mu$ gauge boson; a
scalar of mass $24(g_\mu/\lambda)\tilde m$, which becomes part of the
heavy gauge supermultiplet; and four light scalars with masses
$2\sqrt{6}\tilde m, 2\sqrt{6}\tilde m, 2\sqrt{3}\tilde m$ and
$2\sqrt{2}\tilde m$, correspondingly~\cite{CDM}.

Assuming $\kappa'=0$ for the moment, $\langle S \rangle$ and $\langle
F_S \rangle$ provide the $\mu$ and $B$ terms for the Higgs sector: \be
\mu (M_\mu)= \kappa \langle S \rangle  \simeq 2\sqrt{3} {\kappa \over
\lambda}\tilde m \quad (\gae \tilde m),
\label{mu}
\ee \be B (M_\mu)= {\langle F_S\rangle  \over \langle S  \rangle }
\simeq -2\sqrt{3} {\epsilon\over\lambda} \tilde m  \quad(|B|\ll \tilde
m).
\label{B}
\ee

Below the messenger scale \be M\ \equiv\ f \langle X\rangle \simeq
2\sqrt{3}{\epsilon f\over\lambda^2}\tilde m \quad (\gg \m),
\label{Mql}
\ee the messengers are integrated out, giving rise to the  usual
one-loop gauge mediation contributions to the gaugino masses: \be
M_n(M) = \Delta\beta_n{\alpha_n\over4\pi}\Lambda g\left(\Lambda/
M\right), \ee where $n = 1,2,3$ corresponds to $U(1)_Y$, $SU(2)_W$ and
$SU(3)_C$, $g(x)$ is the threshold function from \cite{Martin} and \be
\Lambda \equiv {\langle F_X \rangle \over \langle X\rangle} \simeq
2\sqrt{3}{\lambda\over\epsilon}\tilde m ~.
\label{Lambda}
\ee The scalar squared masses receive a $U(1)_\mu$ $D$-term
contribution and a negative contribution from the $U(1)_\mu$
mediation: \be m_{\tilde f}^2(M_\mu)  \ =\  z_{f} (4-z_{f}) \tilde m^2,
\label{scalarmass}
\ee in addition to the usual two-loop SM gauge mediation
contributions:  \be m^2_{\tilde f}(M)={2\Lambda^2\over(4\pi)^2} \left(
\Delta\beta_3 C_3^{f}\alpha_3^2 +\Delta\beta_2 C_2^{f}\alpha_2^2
+\frac{5}{3}\Delta\beta_1 C_1^{f}\alpha_1^2 \right)
f\left(\Lambda/M\right) ~,
\label{msq}
\ee where the coefficients $C_i^{f}$ are zero for gauge singlet
sfermions $\tilde f$, and $4/3$, $3/4$ and $y^2$ for fundamental
representations  of $SU(3)_C$, $SU(2)_W$ and $U(1)_Y$,
correspondingly.  The threshold function $f(x)$ can be found in
Ref.~\cite{Martin}.

After imposing electroweak symmetry breaking, the parameter space of
this class of models is spanned by $\{\Lambda, M, M_\mu, \tan\beta,
{\rm sign}(\mu)\}$.  However, if we allow a small coupling $\kappa' X
H_u H_d$, the conditions (\ref{mu}) and (\ref{B}) can be relaxed: \be
\mu (M_\mu)= \kappa \langle S \rangle + \kappa' \langle X \rangle
\simeq 2\sqrt{3}  \left({\kappa \over \lambda}+{\kappa'\epsilon \over
\lambda^2} \right) \tilde m \quad (\gae \tilde m),
\label{mu'}
\ee \be B (M_\mu)= {\kappa \langle F_S\rangle + \kappa' \langle
F_X\rangle \over \kappa \langle S  \rangle + \kappa' \langle X \rangle
}  \simeq -2\sqrt{3}
\left({\epsilon\over\lambda}+{\kappa'\over\kappa}\right) \tilde m
\quad(|B|\lae \tilde m),
\label{B'}
\ee so that $\tilde m$ can be traded for $\kappa'/\lambda$ and treated
as an additional free parameter.  This is particularly relevant for
models with $z_{H_d} < z_{H_u}$, where it is rather difficult to
obtain proper electroweak symmetry breaking at large values of
$\tan\beta$, which are suggested by (\ref{B}).  This can be easily
understood as follows.  Minimization of the tree-level potential leads
to the approximate relation \be m^2_{H_d}(M_Z)-m^2_{H_u}(M_Z) \simeq
m_A^2(M_Z) \ee which implies that $m^2_{H_d}(M_Z) > m^2_{H_u}(M_Z)$.
{}From eq.~(\ref{scalarmass}), however, one finds  \be
m^2_{H_d}(M_\mu)-m^2_{H_u}(M_\mu) = 8 (z_{H_d}- z_{H_u}) \tilde m^2,
\ee so that at the $U(1)_\mu$-breaking scale we already have
$m^2_{H_d}(M_\mu)<m^2_{H_u}(M_\mu)$. In addition, at large $\tan\beta$
the bottom and tau Yukawa couplings are enhanced and tend to further
reduce $m^2_{H_d}(M_Z)$.

The collider phenomenology of this class of models depends on the
nature and lifetime of the next-to-lightest supersymmetric particle
(NLSP). Note that our models have automatic conservation of
$R$-parity, which  can be defined as (recall that we are using the
normalization $z_S=4$)  \be  R\ =\ (-1)^{3[z-6y(z_Q-1)]+2s},  \ee
where $y$ and $z$ stand for the hypercharge and $U(1)_\mu$ charge of a
particle, and $s$ is its spin. Therefore, the NLSP can only decay to
its superpartner plus a gravitino $\tilde G$.

First we discuss the mass spectrum, in order to determine which
particles are potential NLSP candidates. Below the scale $M_\mu$ there
are 6 neutralinos, for which we choose the basis $\{ \tilde B$
,$\tilde W_3$, $\tilde H_d$, $\tilde H_u$,
$\tilde\Sigma\equiv\cos\theta\tilde N+\sin\theta\tilde S'$,  $\tilde
X'\}$, where $\cos^2\theta\approx 2/3$ and
$$
\tilde S'\ = \ {\lambda \tilde S + \epsilon \tilde X \over
\sqrt{\lambda^2+\epsilon^2}}, \qquad\qquad \tilde X'\ = \ {\lambda
\tilde X - \epsilon \tilde S \over \sqrt{\lambda^2+\epsilon^2}}.
$$
Here $\tilde N$ ($\tilde X$, $\tilde S$) denotes the fermionic
component of the SM singlet superfield $N$ ($X$, $S$).  The neutralino
mass matrix is given by \be {\cal M}_{\tilde \chi^0}\ =\ \left(
\ba{cccccc} M_1 & 0   & -{1\over2}g'v_d & {1\over2}g'v_u &     0 & 0
\\ [2mm] 0   & M_2 & {1\over2}gv_d &-{1\over2}gv_u     &     0 & 0 \\
[2mm] -{1\over2}g'v_d     & {1\over2}gv_d & 0  &-\mu
&-{1\over\sqrt{6}}\kappa v_u      &-{1\over\sqrt{2}}\kappa' v_u \\
[2mm] {1\over2}g'v_u & -{1\over2}gv_u &-\mu& 0
&-{1\over\sqrt{6}}\kappa v_d     &-{1\over\sqrt{2}}\kappa' v_d \\
[2mm] 0   & 0   &-{1\over\sqrt{6}}\kappa v_u &-{1\over\sqrt{6}}\kappa
v_d &     0      & 4\tilde m \\  [2mm] 0   & 0
&-{1\over\sqrt{2}}\kappa' v_u &-{1\over\sqrt{2}}\kappa' v_d & 4\tilde
m  & 0 \ea \right), \ee where $v_{u, d}=\sqrt{2}\langle
H_{u,d}\rangle$.  This situation resembles the next-to-minimal
supersymmetric standard model (NMSSM) \cite{NMSSM}, except that now we
have not one, but two singlet states, which are degenerate to lowest
order.

The neutral Higgs masses are the same as in the MSSM, with the
addition of two new CP-even singlet states with masses
$2\sqrt{6}\tilde m$ and $2\sqrt{2}\tilde m$, and two new CP-odd
singlet states with masses $2\sqrt{6}\tilde m$ and $2\sqrt{3}\tilde
m$.  The mixing between these new states and the Higgses of the MSSM
($h^0$, $H^0$ and $A^0$) is suppressed by the small Yukawa couplings
$\kappa$ or $\kappa'$.

In Table~\ref{spectra} we list sample particle spectra for model
points in each of the cases represented in
Table~\ref{Tablesolutions}. In addition to the values of the model
parameters, for completeness we also give the corresponding ratios of
the fundamental parameters in the Lagrangian (coupling constants).
\begin{table}[h!p]
\centering \renewcommand{\arraystretch}{1.3}
\begin{tabular}{||c||c|c|c|c|c||} \hline\hline
Particle         & \multicolumn{5}{c||}{Models}             \\
                 \cline{2-6} & 1a(i)  & 2a(i) & 2a(ii)& 2a(iii)&
                 2a(iv) \\ \hline\hline $\tilde\chi^0_1$ & 130.8  &
                 164   & 81    & 120    & 120    \\ \hline
                 $\tilde\chi^0_2$ & 202    & 268   & 134   & 120    &
                 120    \\ \hline $\tilde\chi^0_3$ & 400    & 724   &
                 507   & 126.5  & 161    \\ \hline $\tilde\chi^0_4$ &
                 400    & 724   & 509   & 201    & 258    \\ \hline
                 $\tilde\chi^0_5$ & 575    & 793   & 544   & 383    &
                 451    \\ \hline $\tilde\chi^0_6$ & 580    & 797   &
                 544   & 401    & 465    \\ \hline\hline
                 $\tilde\chi^+_1$ & 131.0  & 268   & 134   & 200    &
                 258    \\ \hline $\tilde\chi^+_2$ & 581    & 798   &
                 513   & 401    & 466    \\ \hline\hline $\tilde e_R $
                 & 253    & 262   & 248   & 131    & 155    \\ \hline
                 $\tilde e_L    $ & 247    & 427   & 272   & 217    &
                 266    \\ \hline
$\tilde\tau_1  $ & 166    & 147   & 216   & 125.2  & 125    \\ \hline
$\tilde\tau_2  $ & 312    & 478   & 300   & 220    & 277    \\
\hline\hline
$\tilde g      $ & 1126   & 1141  & 615   & 924    & 1134   \\ \hline
$\tilde t_1    $ & 984    & 1045  & 589   & 795    & 979    \\ \hline
$\tilde u_R    $ & 1074   & 1112  & 610   & 866    & 1061   \\
\hline\hline $h^0           $ & 114    & 113   & 109   & 111    & 114
\\ \hline $H^0           $ & 379    & 487   & 177   & 339    & 454 \\
\hline\hline
$M [{\rm TeV}] $ & 500    & 200   & 100   & 200    & 200    \\ \hline
$\Lambda [{\rm TeV}] $  &  50    & 50    & 25    &  40    & 50     \\
\hline $M_\mu [{\rm TeV}]$  & 10,000 & 1,000 &10,000 & 5,000  & 2,000
\\ \hline $\tan\beta     $ &   35   & 60    & 25    & 10     & 25 \\
\hline $\mu(M_\mu)    $ &  602   & 862   &$-537$ & 387    & 460 \\
\hline $\tilde m      $ &  100   & 182   &  156  & 30     & 30 \\
\hline\hline $\kol          $ & 1.74   & 1.37  & 1.14  & 3.72   & 4.43
\\ \hline $\eol          $ & 0.0069 & 0.0126& 0.0188& 0.0026 & 0.002
\\ \hline $\kpol         $ & 0.0796 &  ---  & ---   & 1.545  & 0.713
\\ \hline\hline
\end{tabular}
\caption{ Sample particle spectra for the models in
Table~\ref{Tablesolutions}.}
\label{spectra}
\end{table}
A few comments are in order at this point.  As we mentioned earlier in
this Section, models with $z_{H_d}<z_{H_u}$ (1a(i), 2a(iii) and
2a(iv)) typically require the presence of the additional coupling
$\kappa'$, in which case $\tilde m$ is an input.  Otherwise, $\tilde
m$ is computed from eqs.~(\ref{B}) and (\ref{Lambda}):   \be   \tilde
m\ =\ \sqrt{{|B\Lambda|\over12}}.    \ee If $\tilde m$ is large, the
usual hierarchy between the left-handed and the right-handed sleptons
may be affected, due to the $U(1)_\mu$ contributions in
eq.~(\ref{scalarmass}). For example, in model 1a(i), where $z_E>z_L$
and $\tilde m$ is sizable, we find $m_{\tilde e_R}>m_{\tilde e_L}$,
contrary to the prediction of the minimal models \cite{dnns,DNS}. In
principle, this inverse slepton mass hierarchy is also possible for
models 2a(iii) and 2a(iv).  This contribution, however, is not
important for the squarks, where the SM gauge-mediated contributions
dominate.  We also find that the $\mu$ parameter is typically larger
than in the minimal gauge-mediated models, due to the negative
$U(1)_\mu$ contributions to $m_{H_u}^2$.  Note the presence of the two
extra degenerate neutralinos in the spectrum. However, because of
their very small couplings, their impact on phenomenology is
negligible, unless one of them is the NLSP -- see the examples for
models 2a(iii) and 2a(iv).

Table~\ref{spectra} rather nicely illustrates all potential NLSP
candidates in our models:
\begin{enumerate}
\item 
The lightest neutralino, which is mostly wino-like:
$\tilde\chi^0_1\sim \tilde W_3^0$.  This situation may arise in any
one of the models with $\Delta\beta_2=1$, where at the {\em weak}
scale we find the reversed gaugino mass hierarchy
$M_3:M_1:M_2\sim9:1.5:1$. Since $M_2$ is also the soft mass of the
wino-like chargino,  one faces the dilemma of deciding which one is
actually the NLSP: the chargino or the neutralino.  (Quite recently,
the case of $M_2<M_1$ was discussed in the framework of
supergravity-mediated (SUGRA) models, where the soft masses arise
through the super-conformal anomaly \cite{Randall,GLMR}.)  At
tree-level, one can expand the lightest chargino and neutralino mass
eigenvalues in terms of $1/|\mu|$: \bear m_{\tilde\chi^+_1}&=&M_2
-{M_W^2\over\mu  }s_{2\beta} -{M_W^2\over\mu^2}M_2 + {\cal
O}({1\over\mu^3}), \\ m_{\tilde\chi^0_1}&=&M_2 -{M_W^2\over\mu
}s_{2\beta}           -{M_W^2\over\mu^2}M_2
-{M_W^2\over\mu^2}{M_W^2\over M_1-M_2}t_W^2s_{2\beta}^2 + {\cal
O}({1\over\mu^3}), \eear where $t_W\equiv\tan\theta_W$ and
$s_{2\beta}\equiv\sin 2\beta$.\footnote{Our result for both the
chargino and neutralino differs from that of
Refs.~\cite{Randall,GLMR}.}  We find that the mass splitting occurs
only at order $1/|\mu^2|$ and the chargino is always heavier at
tree-level: \be \Delta m_\chi\ \equiv \
m_{\tilde\chi^+_1}-m_{\tilde\chi^0_1} \ = \  {M_W^2\over \mu^2}
{M_W^2\over M_1-M_2}t^2_W s^2_{2\beta} + {\cal O}({1\over\mu^3}).
\label{delta_mchi}
\ee Notice the additional suppression at large $\tan\beta$ due to the
factor $\sin^22\beta \sim 4/\tan^2\beta$, in which case the next order
terms may be numerically important as well.  Typical values of the
parameters result in a mass splitting $\Delta m_\chi$ in the MeV
range. In any case, we see that in order to correctly determine the
nature of the NLSP, it is necessary to account for the one-loop
gaugino mass corrections \cite{inomasses}.  Including the full
one-loop mass corrections to the chargino and neutralino matrices
\cite{BMPZ}, we find that the neutralino is indeed the NLSP, and the
mass splitting is in fact much larger than predicted by
eq.~(\ref{delta_mchi}). We illustrate this result in
Fig.~\ref{delta_m}.
\begin{figure}[ht]
\epsfysize=3.5in \epsffile[-40 220 320 600]{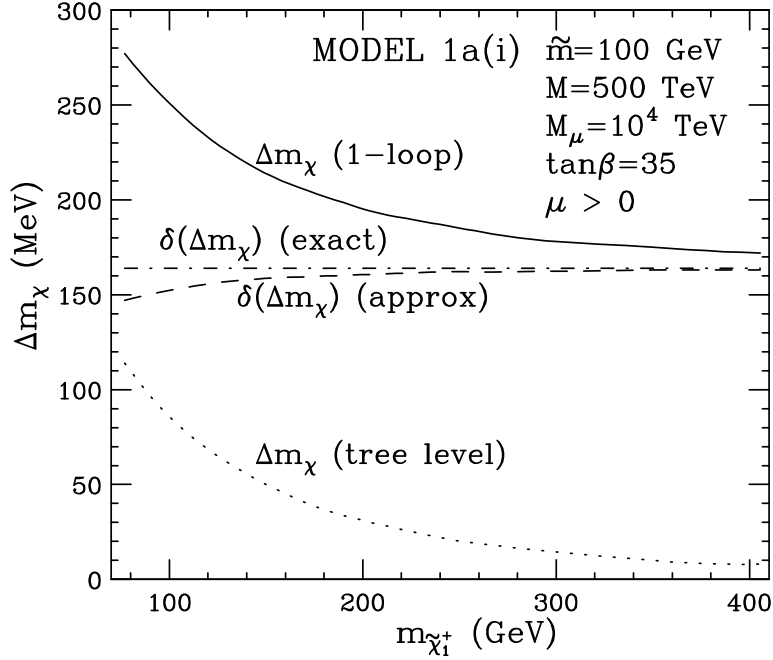}
\begin{center}
\parbox{5.5in}{
\caption[] {\small The mass splitting $\Delta m_\chi\equiv
m_{\tilde\chi^+_1}-m_{\tilde\chi^0_1}$ at tree-level (dotted) and
one-loop (solid), versus the chargino mass $m_{\tilde\chi^+_1}$, which
is varied by varying $\Lambda$.  The dot-dashed line represents the
exact one-loop correction $\delta\Delta m_\chi$, and the dashed line
is the result from the approximation in eq.~(\ref{approximation}).
\label{delta_m}}}
\end{center}
\end{figure}
Even though the chargino and neutralino mass corrections themselves
are dominated by the squark and Higgs loops,  we have checked that the
renormalization of the {\em mass splitting} is due almost entirely to
the gauge boson loops. For small chargino or neutralino mixing, and
keeping only the gauge boson contributions, we can derive the
following approximate formula for the one-loop correction to $\Delta
m_\chi$: \bear \delta\Delta m_\chi &\equiv& \Delta
m_\chi^{1-loop}-\Delta m_\chi^{tree}  \nonumber \\
&=&{g^2\over8\pi^2}
\biggl[2c_W^2B_0(M_2,M_2,M_Z)+2s_W^2B_0(M_2,M_2,0)-2B_0(M_2,M_2,M_W)
\nonumber \\
&-&c_W^2B_1(M_2,M_2,M_Z)-s_W^2B_1(M_2,M_2,0)+B_1(M_2,M_2,M_W)
\biggr]M_2,
\label{approximation}
\eear with the functions $B_0$ and $B_1$ defined as in Appendix B of
Ref.~\cite{BMPZ}. Notice that this correction is purely finite and
cannot be accounted for in a leading-log decoupling scheme.  Since the
dominant effect is from the gauge boson loops only, the result
(\ref{approximation}) is quite model-independent and will apply for
the supergravity-mediated models discussed in
Refs.~\cite{Randall,GLMR} as well.

Since the lightest chargino and neutralino are so degenerate, the
decay length $L_{\tilde\chi}$ for the decay
$\tilde\chi^+_1\rightarrow\tilde\chi^0_1+X$ could be macroscopic
\cite{CheDreGun}: \be L_{\tilde\chi}\ =\  \left({1 GeV\over\Delta
m_\chi}\right)^5 \left( {E^2\over m_{\tilde\chi}^2}-1\right)^{1/2}
\times 100\, \mu m.  \ee
For typical mass splittings $\Delta m_\chi\sim 200$ MeV (see
Fig.~\ref{delta_m}), $L_{\tilde\chi}$ is on the order of tens of
centimeters.  In that case, the lightest chargino and neutralino may
act as co-NLSP's, if the decays to gravitinos are faster.

\item In any one of our models, the limit $\tilde m\rightarrow 0$
gives rise to a neutralino NLSP, which is a mixture of $\tilde\Sigma$
and $\tilde X'$. We see such examples in Table~\ref{spectra} for
models 2a(iii) and 2a(iv), but we find that small $\tilde m$ is
possible for all other models as well.

\item In all models with $\Delta\beta_2=2,3$ or 4 we find that
$M_1<M_2$, so that the lightest neutralino is mostly $\tilde B$, as in
the conventional SUGRA or minimal gauge-mediated models.  For either
moderate values of $\tan\beta$ or rather large values of $\tilde m$,
it also turns out to be the NLSP -- see e.g. model 2a(ii) in
Table~\ref{spectra}.  The phenomenology of similar gauge-mediated
models, albeit with a somewhat different gaugino mass splitting, has
been extensively discussed in the literature \cite{BinoNLSP}.
\item The lightest tau slepton $\tilde\tau_1$ can be the NLSP if
$\tan\beta$ is significant and $\tilde m$ is not too large, e.g. in
model 2a(i) of Table~\ref{spectra}.  This case is not much different
from the minimal gauge-mediated models with a stau NLSP and has been
studied previously \cite{FM,stauNLSP} for both stable and promptly
decaying staus.
\end{enumerate}

The other important factor in the discussion of the typical collider
signatures of our models is the value of the intrinsic SUSY breaking
scale $E_{\rm vac}$, which determines the decay length $L_{\rm NLSP}$
of the corresponding NLSP: \be L_{\rm NLSP}\ \sim\  130 \left( {100\
{\rm GeV}\over m_{\rm NLSP}} \right)^5 \left( {E_{\rm vac}\over 100\
{\rm TeV} } \right)^4 \mu m, \ee with $E_{\rm vac}^4$ being the vacuum
energy density.  The value of $E_{\rm vac}$ in our models is given
by~\cite{CDM}  \be  E_{\rm vac}\ \gae \ {\cal O}(1)\ \left({4\pi\over
g_\mu}  \right) \sqrt{F_X}\ \gae \ {\cal O}(1)\times 200 {\rm TeV}.
\ee  We see that for $E_{\rm vac}$ close to the lower limit ($\sim
10^5$ GeV), $L_{\rm NLSP}$ could be microscopic and unlike most known
models of  gauge-mediated SUSY breaking, prompt decays of the NLSP are
possible.

In the rest of this section we shall concentrate on the first two NLSP
options, since they are unique features of our models. {\em Prompt}
decays of the $\tilde W$-like chargino and $\tilde W_3\,$-like
neutralino co-NLSP's in the models with $\Delta\beta_2=1$ lead to
signatures which have never before been discussed as possible SUSY
discovery modes, so we devote the next subsection \ref{subsec:su2} to
this case.  Later in subsection~\ref{subsec:singletino} we discuss the
phenomenology of the singletino NLSP scenario, which resembles
somewhat that of a gauge-mediated NMSSM.  Finally, we conclude this
Section with comments on the more standard cases of $\tilde B$-like
neutralino or stau NLSP.

\subsection{$SU(2)$-neutralino NLSP}
\label{subsec:su2}

Type 1 models (see Table~\ref{TableMess}) have the generic prediction
$M_2<M_1<M_3$ and the lightest neutralino is mostly  $\tilde W_3$. As
shown in the previous subsection, the lightest  neutralino and the
lightest chargino in this case are degenerate enough so that they can
act as co-NLSP's.  The typical experimental signatures therefore
depend on which chargino-neutralino combinations are mostly being
produced.  At the Tevatron, the dominant production processes are
$p\bar{p}\rightarrow\tilde\chi^+_1\tilde\chi^-_1$ and
$p\bar{p}\rightarrow\tilde\chi^\pm_1\tilde\chi^0_1$, which are roughly
of the same order, while
$p\bar{p}\rightarrow\tilde\chi^0_1\tilde\chi^0_1$ is much smaller. In
the rest of this subsection, we shall therefore only consider the fate
of a $\tilde\chi^+_1\tilde\chi^-_1$ or a
$\tilde\chi^\pm_1\tilde\chi^0_1$ gaugino pair.

If the SUSY breaking scale $E_{\rm vac}$ is high, the decays of both
the chargino and the neutralino to gravitinos will happen outside the
detector and the signatures are similar to those discussed in
Ref.~\cite{CheDreGun,Randall,GLMR} for supergravity-mediated models.
In this case the chargino will have time to decay to a neutralino
first.  However, it is rather unlikely that the chargino will make it
out to the muon chambers -- we saw that the one-loop corrections tend
to increase the $\tilde\chi^\pm_1-\tilde\chi^0_1$ mass splitting and
the chargino decay will probably occur within a meter or so from the
primary vertex, thus evading existing limits from heavy charged stable
particle searches \cite{CDF-heavy}.  It will therefore look like a tau
and will be rather difficult to identify \cite{Randall}.  Because of
the small chargino-neutralino mass splitting, the lepton from the
$\tilde\chi^\pm_1\rightarrow\tilde\chi^0_1 l^\pm \nu$ decay will be
very soft and cannot be used to tag the chargino decay.  Note also
that this mass degeneracy renders the current LEP limits on the
chargino mass inapplicable.

As in any model with a rather low SUSY breaking scale, decays of the
NLSP to $\tilde G$ provide information about the hidden (or messenger)
sector via $L_{\rm NLSP}$.  If it is finite ($\gae 1$ mm) and the
NLSP's ($\tilde\chi^\pm_1$ or $\tilde\chi^0_1$) decay to gravitinos
inside the detector, this will give rise to events with displaced
vertices (kinks in the charged tracks), photons with finite impact
parameters or originating from the outer hadronic calorimeter
\cite{CheGun}.  A recent CDF search for long-lived $Z$-parents
\cite{CDF-Zparents} is not sensitive enough to place a limit on the
neutralino mass in this case.  Because of the phase space suppression,
the branching ratio $BR(\tilde\chi^0_1\rightarrow Z\tilde G)$ begins
to dominate over $BR(\tilde\chi^0_1\rightarrow \gamma\tilde G)$ only
for neutralino masses $m_{\tilde\chi^0_1}\gae 130$ GeV, where the
production cross-section falls below the Run I sensivity.

Finally, if the SUSY breaking scale $E_{\rm vac} \sim 10^5$ GeV, the
chargino and neutralino co-NLSP's may decay promptly to gravitinos,
creating events with real $W$'s, $Z$'s or photons and missing
(transverse) energy.  Since the signatures for
$\tilde\chi^+_1\tilde\chi^-_1$ and $\tilde\chi^\pm_1\tilde\chi^0_1$
production are different, we shall discuss each case in turn.

For chargino pair production with subsequent prompt decays to
gravitinos, the possible final state signatures are
$l^+l^-\!\not\!\!E_T$, $ljj\!\not\!\!E_T$ and $jjjj\!\not\!\!E_T$,
with branching ratios 6\%, 38\% and 56\%, correspondingly.  The two
leptonic signatures suffer from large irreducible  $W$-pair and
$t$-$\bar{t}$ backgrounds, although the latter one may be somewhat
suppressed via a $b$-jet veto. These two channels have been previously
considered as possible Standard Model Higgs search modes at both the
Tevatron and LHC \cite{Herbi,Han,Andre}, since for $m_h>140$ GeV the
branching ratio $BR(h\rightarrow W^+W^-)$ starts to dominate.  The
result is that this signal will be rather difficult to observe at the
Tevatron, and a $3\sigma$ discovery is only possible with Run III
integrated luminosities $L_{\rm int}\sim 30\, {\rm fb}^{-1}$
\cite{Andre}.  For a certain range of chargino masses,  we can
immediately adapt this result to our case.  For Higgs masses in the
range $140-180$ GeV, the cross-section for $W$-pair production via
single Higgs is $\sigma_{h}(gg\rightarrow h^0\rightarrow WW)\sim
0.2-0.4$ pb.  For chargino masses in the range 130-150 GeV, the signal
cross-section $\sigma_{\tilde\chi}(p\bar{p}\rightarrow \chi^+\chi^-
+X)$ is of the same order, so we conclude that only Run III at the
Tevatron may possibly have some sensitivity beyond LEP-II in those two
channels.  For smaller chargino masses, the Tevatron reach is better
and a signal may be observed in the very early stages of Run III.  In
the most optimistic scenario, where the chargino mass is just beyond
the projected LEP-II limit ($m_{\tilde\chi^+_1}\sim 100$ GeV),
$\sigma_{\tilde\chi}\sim 1.2$ pb and can be observed even in Run II.

The other possible signal of  $\tilde\chi^+_1\tilde\chi^-_1\rightarrow
W^+W^-\tilde G\tilde G$ is the multijet channel, which has rather
small SM physical backgrounds  (the $t$-$\bar{t}$ background can be
suppressed with a lepton veto). The single Higgs production analogy
now does not work, because of the $\not\!\!\!E_T$ requirement.  The
dominant background is from QCD multijet production and jet energy
mismeasurement, which is why a detailed Monte Carlo study with a very
realistic detector simulation is necessary in order to estimate the
reach in this channel. In addition to a hard $\not\!\!E_T$ cut, one
may also make use of the fact that two different jet pairs should
reconstruct close to the $W$ mass.

We now turn to the signatures arising in the
$\tilde\chi^\pm_1\tilde\chi^0_1$ case, where we have to factor in the
branching ratios of the neutralino to a $Z$ or a photon.  For
relatively light neutralinos, it is best to study signatures where the
neutralino decays to a photon and a gravitino.  First, for
$m_{\tilde\chi^0_1}\lae M_Z$, this is the dominant decay mode ($\sim
100\%$) anyways. Second, even when $m_{\tilde\chi^0_1} > M_Z$ and the
decay to $Z$ dominates, the $BR(\tilde\chi^0_1\rightarrow \gamma
\tilde G)$ is never below $\sim 20\%$, which is still better than the
leptonic branchings of the $Z$'s (the channels with hadronic $Z$'s
have larger backgrounds). We conclude that the most promising clean
signature in this case is $l^\pm\gamma\not\!\!\!E_T$. The only
physical background process is $W\gamma$, which is rather rare, so the
typical backgrounds will involve photon/lepton misidentification
and/or $E_T$ mismeasurement.  Note that in contrast to the minimal
gauge-mediated models,  our type 1 models are not associated with any
{\em di-photon} signatures \cite{diphoton}, because the neutralino
pair-production cross-sections are suppressed, while the chargino
decay does not yield a photon.

Finally, there is a variety of possible signatures, if we consider
prompt neutralino decays to $Z$'s. We shall concentrate on the
following channels: $l^+l^- l^\pm\not\!\!\!E_T$; $l^+l^- jj
\not\!\!\!E_T$; $l^\pm  jj   \not\!\!\!E_T$ and $jjjj \not\!\!\!E_T$,
since $l^\pm  \not\!\!\!E_T$ and $jj \not\!\!\!E_T$ have too large a
background to be even considered.

The clean trilepton signature has irreducible background from $WZ$ and
in addition one takes a hit from the $Z$ branching ratio of the
neutralino, so it is rather unlikely that an excess of such events
will be seen in any of the future Tevatron runs.  Unlike the classic
SUSY trilepton signature \cite{3L}, one cannot use an invariant
dilepton mass cut to beat down the $WZ$ background.  The case of the
$l^\pm jj\not\!\!\!E_T$ is even worse: it has large irreducible
backgrounds from both $WZ$ and $t\bar{t}$.

The dilepton plus jets signature $l^+l^-jj\not\!\!E_T$ looks somewhat
promising. It was used to search for cascade decays of gluinos and/or
squarks \cite{gluinosquark}.  The difference now is that the leptons
are coming from a $Z$-decay, so the invariant dilepton mass cut is
exactly the opposite of what is used in the conventional SUSY search.
The dominant physical backgrounds then would be $Zjj\rightarrow
\tau^+\tau^- jj\rightarrow l^+l^- jj\not\!\!E_T$, and to some extent
$t$-$\bar{t}$. Both of them can be significantly reduced by requiring
that the jet pair reconstructs the $W$ mass.

The 4-jet plus $\not\!\!E_T$ signature was already discussed above for
the case of hadronically decaying $W$'s in chargino pair-production,
the difference now is that the two jet pairs should reconstruct the
$W$ and $Z$ mass, correspondingly, so that one should use a more
relaxed cut, e.g., $70\ {\rm GeV}<m_{jj}<100$ GeV.

\subsection{Singletino NLSP}
\label{subsec:singletino}

In the limit of small $\tilde m$ the two lightest neutralinos will be
rather degenerate and have significant ``singletino'' components from
$\tilde \Sigma$ and $\tilde X'$. Since their masses are of order
$4\tilde m$, while the mass of the lightest scalar singlet $H_S$ is
only $2\sqrt{2}\tilde m$, the ``singletino''-like NLSP will always
decay as  $\tilde\chi^0_1\rightarrow H_S\tilde G$. $H_S$ will
subsequently decay to $b$-$\bar{b}$, due to the small
$S$-$\{H_u,H_d\}$ mixing.  If the singletino decays some distance away
from the primary vertex, this will give rise to rather spectacular
signatures with displaced $b$-jets.  The case when the singletinos
decay promptly resembles that of the minimal gauge-mediated models
with a short-lived higgsino NLSP \cite{KTM}, heavier than the light
Higgs $h^0$.  The difference now is that the jet pairs should
reconstruct the mass of the singlet Higgs $H_S$ rather than
$h^0$. Note that the LEP limits on the Higgs mass do not directly
apply to $H_S$.

If the singletinos decay outside the detector, the typical signatures
depend on the nature of the next-to-next-to-lightest supersymmetric
particle\footnote{We do not count the second singletino.} (NNLSP).
Because of the small couplings of the `singletinos', all
supersymmetric particles will first decay to the NNLSP. For the models
from Table~\ref{spectra}, the NNLSP  is typically $\tilde\tau_1$,
which can be understood as follows. The singletino NLSP case arises
for small values of $\tilde m$, when the $U(1)_\mu$ contributions to
the scalar masses are also small. Then, the supersymmetric mass
spectrum in any of our models resembles that of a minimal
gauge-mediated model, with the corresponding number and type of
messenger representations. Thus we can immediately adapt the NLSP
analysis in the minimal gauge mediated models to the  question of the
NNLSP in our models. The balance between the masses of the two main
NNLSP candidates: stau and $\tilde B$-like neutralino, is for the most
part determined by the value of the messenger multiplicity factor
$\Delta\beta_1$, since $m_{\tilde\tau_1}\sim \sqrt{\Delta\beta_1}$,
while $m_{\tilde B}\sim \Delta\beta_1$. In models of type 1 and 2,
$\Delta\beta_1$ is large, and the stau is lighter than the bino
throughout most of the parameter space.  One should keep in mind
though that in models 1 the stau mass should be compared to the
$\tilde W_3$-like neutralino mass instead, so that cases with
$m_{\tilde\chi^0_1}<m_{\tilde W_3}<m_{\tilde \tau_1}<m_{\tilde B}$ are
certainly possible.  Note that at low enough values of $\tan\beta$ and
$\Lambda$ one can reach a situation where
$m_{\tilde\chi^0_3}-m_{\tilde \tau_1}<m_{\tau}$, so that the stau and
the bino are in fact co-NNLSP's. Such an example is shown in
Table~\ref{spectra} for model 2a(iii).  Next, for $\Delta\beta_1=2$
(models of type 4), one typically finds $m_{\tilde\chi^0_3}<m_{\tilde
\tau_1}$.  Finally, for $\Delta\beta_1=3$ (models of type 3), one
finds cases with either stau or bino NNLSP.

Turning on to the collider phenomenology of models with stable
singletino NLSP, we first discuss the stau NNLSP case.  In principle,
each SUSY event will contain at least two taus from the
$\tilde\tau_1\rightarrow \tilde\chi^0_1 \tau $ decays.  Their $p_T$
spectrum is determined by the mass difference
$m_{\tilde\tau_1}-m_{\tilde\chi^0_1}$, and may be quite soft -- see
the 2a(iii) example in Table~\ref{spectra}.  To make matters worse,
the tau jets and especially the leptons from the tau decays will be
even softer, presenting serious triggering and identification problems.

The distinctive collider signature in case of a neutralino co-NNLSP
depends on which is the dominant decay of $\tilde\chi^0_3$ to the
singletino NLSP. There are three possibilities:
\begin{figure}[t!]
\epsfysize=4.0in \epsffile[-50 200 420 690]{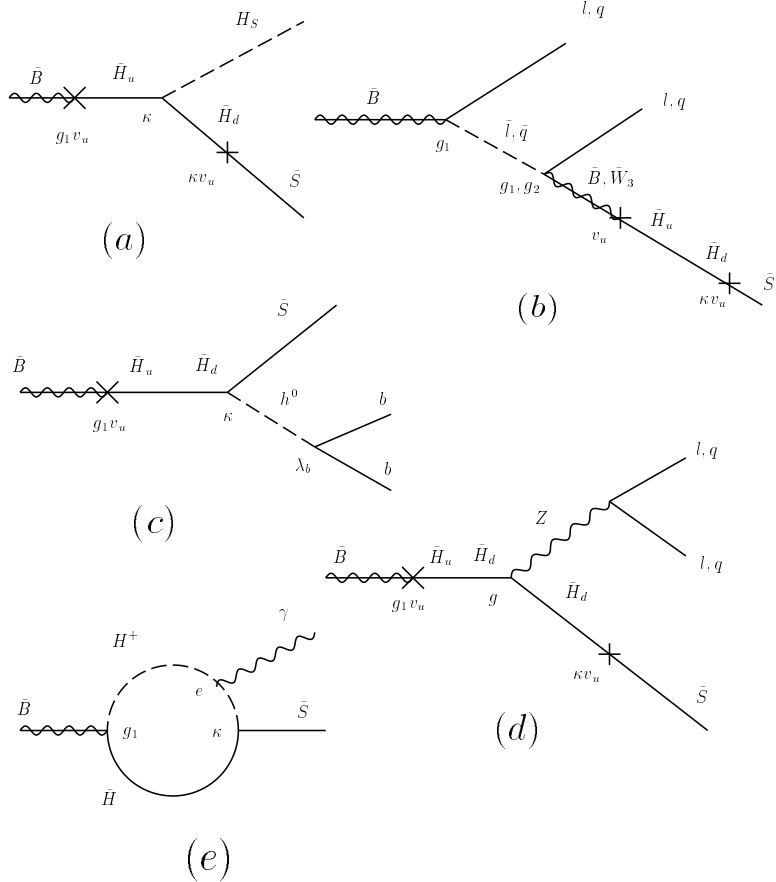}
\begin{center}
\parbox{5.5in}{
\caption[] {\small Sample diagrams for the possible decay modes of
$\tilde\chi^0_3$ to the singletino NLSP $\tilde\chi^0_1$.
\label{diagrams}}}
\end{center}
\end{figure}
\begin{enumerate}
\item The two-body decay $\tilde\chi^0_3\rightarrow\tilde\chi^0_1 H_S$
may be open for $\tilde m \lae M_1/6.8$.  This decay proceeds via the
diagram shown in Fig.~\ref{diagrams}(a) and one can see that the rate
is suppressed by four powers of $\kappa$ or $\kappa'$, as well as the
gaugino-higgsino mixing.
\item For values of $\tilde m \gae M_1/6.8$, the tree-level two body
decays of $\tilde\chi^0_3$ are closed and the three-body decays via
the diagrams in Fig.~\ref{diagrams}(b)-(d) are possible. They are
typically suppressed by only two powers of $\kappa$ or $\kappa'$, in
addition to the gaugino-higgsino mixing.
\item The radiative decay $\tilde\chi^0_3\rightarrow\tilde\chi^0_1
\gamma$ (Fig.~\ref{diagrams}(e)) is also possible. It becomes
important when the $\tilde B$ and $\tilde \Sigma$ ($\tilde X'$) masses
are very close and the three-body decays are suppressed. Unlike the
previous decays, this mode has no gaugino-higgsino mixing suppression.
\end{enumerate}
The relative importance of these three modes will depend on the
particular values of the model parameters \cite{singletino}.  A more
quantitative analysis will have to take into account the correct
singletino-gaugino-higgsino mixing as well as the singlet-Higgs mixing.

We conclude this Section with some comments on the more conventional
cases of $\tilde B$ or stau NLSP. For the most part, they are very
similar to the corresponding minimal gauge-mediated models, and the
results from previous phenomenological analyses hold
\cite{BinoNLSP,FM,stauNLSP}.  However, there are two
differences. First, the predicted gaugino mass ratios are different.
This is important e.g. in the case of a `stable' Bino-NLSP, since the
$p_T$ distributions of the $\tilde\chi^+_1$ and $\tilde\chi^0_2$ decay
products will be affected.  For a given $\tilde\chi^+_1$ mass
(i.e. signal cross-section), we would expect softer (harder) $p_T$
spectra for models 2 (3-4), which will have an impact on the cuts
optimization. Second, in the minimal gauge-mediated models, for
$\mu>0$, large\footnote{The exact numerical bound depends on $\Lambda$
and the number of messenger pairs.}  values of $\tan\beta$  are
typically excluded because the light stau is below the experimental
limit.  In our models, with the possibility of the stau mass to
receive additional positive contributions from the $U(1)_\mu$ D-term,
we find that the large $\tan\beta$ part of the parameter space for
$\mu>0$ can be extended up to $\tan\beta\sim 70$, where either
$m_A^2<0$ or the tau Yukawa coupling diverges below the Planck scale
(the bottom Yukawa coupling is less of a problem, since for $\mu>0$ it
is reduced by the SUSY threshold corrections).

\section{Discussion and Conclusions}
\label{conclusions}
\setcounter{equation}{0}

In this section we discuss how robust our model selection assumptions
are, we summarize the phenomenological signatures, and we comment  on
the general features of the models.  We start with a list the most
notable constraints on model-building and we comment on their
necessity:

\begin{itemize}
\item
Viability of the models even if any global symmetry  (which is not an
accidental result of the gauge structure) is badly violated by Planck
scale physics.

To this end, the models have to be chiral ({\it i.e.} there are no
gauge invariant mass terms), and generic ({\it i.e.}  there are no
gauge invariant and supersymmetric dimension-four  operators with
exceedingly small coefficients in the Lagrangian;  in practice we may
allow dimensionless couplings as small as  the Yukawa coupling of the
electron). Hence, the $\mu$-term is induced  only after a gauge
symmetry is spontaneously broken, while  baryon number conservation is
a consequence of the gauge symmetry.

This constraint is a major motivation for the model-building  effort
presented in Section~2.  So far there is no rigorous proof that the
global symmetries of the  MSSM are violated by Planck scale physics if
they are not protected  by gauge invariance.   However, this may be
the case, and therefore it is important to  search for extensions of
the MSSM which remain viable independent  of the quantum gravitational
effects.

\item
The minimality of the gauge group: SM $\times U(1)_\mu \times$  DSB.

The gauge group has to include the standard  $SU(3)_C \times SU(2)_W
\times U(1)_Y$ and some DSB gauge group  responsible for breaking
supersymmetry.  It is remarkable that the addition of the $U(1)_\mu$
gauge group  is sufficient to prevent the potentially dangerous Planck
scale  effects and to communicate supersymmetry breaking to the MSSM
fields.  In principle, the $U(1)_\mu$ may be replaced by a larger
gauge group,  but in that case it would be harder to cancel the mixed
gauge anomalies.

\item
The cancellation of the mixed SM $\times U(1)_\mu$ anomalies  of the
MSSM fields by the  messenger sector, and of the remnant $U(1)_\mu$
and $U(1)_\mu^3$  anomalies by the DSB sector.

These are nice features of our models because the existence  of the
three sectors (MSSM, messenger and DSB) is required by the
mathematical consistency   of the theory. This is to be contrasted
with the  original gauge mediated supersymmetry breaking models
\cite{dnns,DNS} where the three sectors  are introduced ad-hoc, for
phenomenological reasons.

\item
The quark and lepton masses are generated by the  Yukawa couplings to
the Higgs vevs.

This assumption is convenient but does not help in explaining the
pattern of observed quark and lepton masses. If one allows only some
of the fermions to couple to the Higgs doublets, while inducing  the
other quark and lepton masses using a Frogatt-Nielsen sector, higher
dimensional operators, or other mechanism, then  the $U(1)_\mu$ charge
assignment can be more general so that completely different models may
be constructed. We will not elaborate  further this possibility.

\item
The neutrinos have masses and the mixings involve all three
generations.

As suggested by the solar, atmospheric, and accelerator  neutrino
experiments, we have allowed the most general Yukawa couplings of the
neutrinos to the Higgs. This constraint can be relaxed, for example if
there are enough sterile neutrinos. In that case the lepton $U(1)_\mu$
charges no longer need to be generational independent.  We also assume
that the Majorana masses for the right-handed neutrinos come from
$\langle N \rangle$, which results in automatic $R$-parity
conservation. If right-handed neutrinos obtain their masses from
$\langle S \rangle$, $R$-parity violating operators which violate
lepton number will exist and their couplings have to be quite small.
Of course, even a small $R$-parity violating coupling can allow the
NLSP to decay to jets and/or leptons instead, thus changing the
typical collider signatures correspondingly.

\item
The $U(1)_\mu$ charges of the quarks and leptons are  positive.

This constraint is sufficient to ensure that the squarks and sleptons
do not acquire vevs, but is not necessary.  There could be regions in
the parameter space where the  positive contributions to the squark
and slepton squared-masses  from standard gauge mediation dominate
over the $U(1)_\mu$  $D$ term contribution. In that case negative
$U(1)_\mu$ charges  for the quarks and leptons may be allowed. Squark
and gluino masses are insensitive to this contribution, but it may
affect the slepton spectrum and the question of NLSP. However, even
restricting ourselves to models with positive $U(1)_\mu$ charges for
quarks and leptons, we have  found examples which exhaust all possible
NLSP canditates, so considering negative charges will not give us
anything new as far as phenomenology is concerned.

\item
The set of SM singlet superfields  from the messenger sector is
minimal.

It is possible to find various ways of extending the  messenger
sector. For example, there can be more $X$ fields, with non-zero vevs
for the scalar and $F$-components, which would result in a more
general squark and slepton spectrum.  However, with more singlets, it
is harder to find  a viable minimum.

\item
The $U(1)_\mu$ charges are reasonably simple.

This assumption is necessary only if one wants to be able  to embed
$U(1)_\mu$ in a (``reasonably simple'')  non-Abelian gauge group.

\item
There are no fields with fractional electric charge.

Such fields would be stable and produced in large numbers in the Early
Universe, which is in disagreement with a  wide range of
experiments. This constraint can be avoided if the number of particles
with fractional electric charge has  been dramatically diluted during
a period of inflation that ended  at a temperature below their masses.
 
\item
The messenger fields can decay via dimension-4 operators.

Otherwise the lightest messenger is long lived and its presence  at
the time of nucleosynthesis is ruled out by cosmological observations.
Again, this constraint can be relaxed if the Universe suffered a
period of late inflation. Without this assumption, we find solutions
for other classes of models as well.

\item
The DSB sector does not give rise to more than two composite  chiral
fermions charged under $U(1)_\mu$.

This assumption was made only for simplicity.
\end{itemize}

We point out that the phenomenology of these models  is rather
insensitive to some of the extensions listed above.  For example, the
last three assumptions itemized do not  affect some of the novel
phenomenological features  discussed in Section~3:
\begin{enumerate}
\item Non-standard (yet predictable) gaugino mass ratios.
\item Light singlet fermion and/or scalar states may sometimes be in
the spectrum.
\item The models allow for the intrinsic SUSY breaking scale $E_{\rm
vac}$ to be quite low, on the order of a few times $10^5$ GeV, thus
allowing prompt decays of the NLSP.  Note that other models with the
SUSY breaking scale below $10^6$GeV are known \cite{LSSB}, but their
viability relies on assumptions about noncalculable strong dynamics
\item In certain cases we find new NLSP candidates: $\tilde W$-like
chargino, $\tilde W_3\,$-like  neutralino or $\tilde S$-like
neutralino (``singletino'').
\end{enumerate}

It is worth emphasizing that the new SUSY discovery signatures of
$WW\not\!\!\!E_T$, $W\gamma\not\!\!\!E_T$ and $WZ\not\!\!\!E_T$ depend
only on two assumptions: $M_2<M_1,M_3$ and a low SUSY breaking scale.
Therefore, the importance of these signatures, which have been
overlooked  until now, transcends the models introduced in this paper.
Even though we only discussed the phenomenological signatures of our
models for the case of the Tevatron, it is clear that the LHC, where
statistics is not an issue, will be able to definitively explore these
models via the clean signatures considered in
Section~\ref{phenomenology}.

In conclusion, we have constructed several classes of gauge-mediated
models which provide a rather complete answer to the question of
SUSY-breaking and communication to the visible sector.  The models
allow acceptable neutrino masses, and at the same time avoid the $\mu$
problem and the difficulties with FCNC, baryon number violation and
messenger dark matter.

In retrospect, our models still leave several unsolved puzzles.  Most
importantly, we have not attempted to explain the pattern of quark and
lepton masses. Some relatively small Yukawa couplings are still needed
for them and also for the $U(1)_\mu$ breaking sector.  In addition, we
have not addressed the related strong CP problem, whose solution in
this approach should also follow from some gauge  symmetry. Otherwise,
it would be highly sensitive to Planck scale physics too, as is, for
example, the Peccei-Quinn solution~\cite{strongcp}.  Another open
question is whether the gauge couplings and gauge groups  may unify at
some high scale.  Finally, the vacuum in our model is  metastable
(with a lifetime longer than the age of the
universe~\cite{DDR}), and this raises the question why it was chosen by the
early universe.

\vspace*{1.5cm}

{\it Acknowledgements:} We would like to thank M.~Luty and
S.~Willenbrock for discussions.  Fermilab is
operated by the URA under DOE contract DE-AC02-76CH03000.

\section*{Appendix A: \ The 4-3 model}
\label{43model}
\renewcommand{\theequation}{A.\arabic{equation}}
\setcounter{equation}{0}

The detailed discussion of SUSY breaking in the $SU(4)\times SU(3)$
model can be found in Refs.~\cite{PST,AMM,CDM}. Here we just present
the model with the $U(1)_\mu$ charge assignment,  and a brief
description of the essential results.  The field content and the
$U(1)_\mu$ charges are shown in Table~\ref{tab:dsb}.
\begin{table}[ht]
\centering \renewcommand{\arraystretch}{1.4}
\begin{tabular}{|c||c|c||c|}\hline
Fields & $SU(4)$ & $SU(3)$  & $U(1)_\mu$\\ \hline \hline ${\cal Q}$ &
	  4 & 3 & $-(z_{b_1}+z_{b_2}) / 12$ \\ \hline ${\cal L}_1$ &
	  ${\overline 4}$ & 1  & $(3z_{b_1}-z_{b_2})/ 4$ \\ \hline
	  ${\cal L}_2$ & ${\overline 4}$ & 1  & $(3z_{b_2}-z_{b_1})/
	  4$ \\ \hline ${\cal L}_3$ & ${\overline 4}$ & 1  &
	  $-(z_{b_1}+z_{b_2})/ 4$ \\ \hline ${\cal R}_1$ & 1 &
	  ${\overline 3}$ & $(-2z_{b_1}+z_{b_1})/ 3$ \\ \hline ${\cal
	  R}_2$ & 1 & ${\overline 3}$ & $(z_{b_1}-2z_{b_2})/ 3$ \\
	  \hline ${\cal R}_3, {\cal R}_4$ & 1 & ${\overline 3}$ &
	  $(z_{b_1}+z_{b_2})/ 3$ \\ \hline
\end{tabular}
\parbox{5.5in}{
\caption{Particle content and charge assignments in the DSB sector.
\label{tab:dsb}}}
\end{table}

The superpotential of the DSB sector is given by
\be W_{DSB}\ =\  \lambda_1\eL_1\Q\R_1 + \lambda_2\eL_2\Q\R_2 +
\lambda_3\eL_3\Q\R_3 + {\alpha\over3!}\R_1\R_2\R_4 .  \ee
We assume that $\alpha \ll \lambda_1, \; \lambda_2, \; \lambda_3 \;
\sim 1$, so that the vacuum lies in the weakly coupled regime and
hence calculable.  The low energy degrees of freedom can be described
by the baryons $b_i$, where
\be b_i = \frac{1}{3!} \epsilon_{ijkl} \R_j \R_k \R_l , \ee
with $U(1)_\mu$ charges $z_{b_1},\, z_{b_2}, \, 0,\, 0$, respectively.

The $b_3$ and $b_4$ fields get vevs of the order
$(\alpha^{-\frac{4}{9}} \Lambda_D)^3$, where $\Lambda_D$ represent the
$SU(4)$ scale.The energy density at the minimum and the masses of the
scalar components of $b_1, \; b_2$ are
\bear E^4_{\rm vac} &\sim& \alpha^{\frac{2}{9}} \Lambda_D^4 ,\\
m_{b_{1,2}}^2 \equiv m_b^2 &\sim& \alpha^{\frac{10}{9}} \Lambda_D^2 .
\label{mb}
\eear
At one loop, the $b_1$ and $b_2$ fields will generate a
Fayet-Illiopoulos $D$ term for the $U(1)_\mu$ gauge group,
\be -\xi^2 = - \sum_{j=1,2} \frac{g_\mu^2}{16\pi^2} z_{b_j} m_{b_j}^2
\ln \frac{M_V^2}{p^2} ,
\label{D-term}
\ee
where $M_V$ represents the mass scale of the heavy fields in the DSB
sector, and the lower cutoff scale $p^2$ is the  larger one between
the $U(1)_\mu$ breaking scale, $M_\mu^2$, and $m_b^2$.  They also
generate a negative contribution to the mass squared of each scalar
field charged under $U(1)_\mu$ at two-loop, proportional to the
field's charge squared,
\be \frac{m_i^2}{z_i^2} \equiv -\m^2 = -\sum_{j=1,2} 4 \left(
\frac{g_\mu^2}{16\pi^2}\right)^2 \left(z_{b_j}\right)^2  m_{b_j}^2 \ln
\frac{M_V^2}{p^2} .
\label{2loopmass}
\ee
Note that the formulae (\ref{D-term}), (\ref{2loopmass}) only apply
when $p^2 < M_V^2$. If the $U(1)_\mu$ breaking scale ($p^2=M_\mu^2$)
is higher than $M_V^2$, the results will be suppressed by a factor
$M_V^2/M_\mu^2$.


\vfill
\end{document}